# Tuning of dipolar interactions and evaporative cooling in a three-dimensional molecular quantum gas

Jun-Ru Li[1,*], William G. Tobias[1], Kyle Matsuda[1], Calder Miller[1], Giacomo Valtolina[1], Luigi De Marco[1], Reuben R. W. Wang[1], Lucas Lassablière[2], Goulven Quéméner[2], John L. Bohn[1], and Jun Ye[1,*]

1. JILA, National Institute of Standards and Technology and Department of Physics, University of Colorado, Boulder, CO, 80309, USA

2. Université Paris-Saclay, CNRS, Laboratoire Aimé Cotton, 91405 Orsay, France

## ABSTRACT

Ultracold polar molecules possess long-range, anisotropic, and tunable dipolar interactions, providing unique opportunities to probe novel quantum phenomena. However, experimental progress has been hindered by the predominance of two-body loss over elastic interactions, which also limits further cooling via evaporation. Though recent work has demonstrated controlled interactions by confining molecules to a two-dimensional geometry, a general approach for tuning molecular interactions in a three-dimensional (3D), stable system has been lacking. Here, we demonstrate tunable elastic dipolar interactions in a bulk gas of ultracold $^{40}K^{87}Rb$ molecules in 3D, facilitated by an electric field-induced shielding resonance which suppresses the reactive loss by a factor of 30. The favorable ratio of elastic to inelastic collisions enables direct thermalization, the rate of which depends on the angle between the collisional axis and the dipole orientation controlled by an external electric field. This is a direct manifestation of the anisotropic dipolar interaction. We further achieve dipolar interaction-mediated evaporative cooling in 3D. This work demonstrates full control of a long-lived bulk quantum gas system with tunable long-range interactions, paving the way for the study of collective quantum many-body physics.

## INTRODUCTION

The study of atomic quantum gases has benefited from precise control over interactions between their constituents. By tuning the interactions with convenient tools such as external fields[1,2], one can vary the properties of a quantum system and explore its dynamics and phase transitions. Compared to atoms, polar molecules[3–6] possess large electric dipole moments and rich energy level structures, making them a unique platform for studying a range of topics such as quantum magnetism[7–10], exotic superfluidity[11–13], dipolar collective dynamics[14–19], precision measurement[20,21], quantum sensing[22], and quantum information processing[23,24].

Elastic collisions play a critical role in many of these applications. Yet so far, direct observation of the elastic collisions in 3D ultracold molecular gases has been prevented by rapid collisional losses[25–29], even for molecular species without exothermic chemical reactions[27–29], although the mechanisms responsible are still under investigation[30–34]. The recent production of a degenerate Fermi gas of $^{40}$K$^{87}$Rb (KRb) molecules led to the surprising discovery of suppressed reaction rates upon entering deep degeneracy[35,36], though the effect was observed at zero field where the dipolar elastic collisions vanish.

Turning on the dipolar interaction in a 3D geometry by applying an external electric field $E$ to polarize the molecules leads to vastly enhanced losses owing to dipolar attraction[26,37]. While the collision cross section of dipolar elastic collisions increases with the induced dipole moment $d$ as $d^4$ (Ref.[38]), reactive losses increase more strongly as $d^6$ (Ref.[26,37]) for fermionic molecules, preventing the observation of any dynamics related to the elastic dipolar collisions. Inspired by earlier theory[39,40] and experimental work[41], we recently demonstrated that in quasi-two-dimensions (quasi-2D)[42], where the geometry restricts the relative angles of colliding molecules, the two-body loss can be suppressed by more than a factor of 2 from the zero-field value and strong elastic collisions dominate.

Despite the progress made in quasi-2D, a full three-dimensional (3D) gas of polar molecules provides the most general platform for studying dipolar physics. Moreover, interaction effects associated with the long-range and anisotropic nature of the dipolar interaction are more prominent in 3D[43], leading to unique collective dynamics in bulk dipolar gases[14–18]. Therefore, the ability to tune the dipolar interaction in 3D while mitigating the strong two-body loss is of great experimental interest. Resonant shielding, recently demonstrated in quasi-2D[44–46], is a promising approach for suppressing molecular loss in 3D. However, the properties of elastic collisions under the shielding were not experimentally studied.

In this work, we use resonant shielding to enable full control over the interactions between ultracold, reactive molecules in 3D, accessing a new regime where elastic collisions are dominant. At the shielding field of $|E_s|$ = 12.72 kV/cm, we observe a suppression of the two-body loss rate between molecules in the first excited rotational state by a factor of 30 below the background value and achieve a lifetime of ~10 s at typical densities $n$ of 2.5 × $10^{11}$ cm$^{-3}$. Enabled by the shielding, we quantitatively characterize elastic collisions in the system by performing cross-dimensional relaxation experiments after selectively heating along one trap axis of the molecular gas. An elastic-to-inelastic collision ratio $\gamma$ of 12 is measured. We further show that the relaxation rate changes by a factor of 2.5 as we vary the orientation of the dipoles relative to the direction of heating. Despite the presence of the shielding, our measurements of the elastic collisions are consistent with universal dipolar scattering[38,47]. Leveraging the large $\gamma$ at $\boldsymbol{E}_S$, we perform efficient evaporative cooling in 3D, signaled by a gain in the phase-space density (PSD) as particle number is reduced. Since the shielding mechanism is predicted to work for a broad class of molecular species[48], our strategy provides a general method for preparing low-entropy molecular samples with strong elastic dipolar interactions.

Our experiments are conducted with KRb in the excited rotational state $|N = 1, m_N = 0\rangle$ in a 3D optical dipole trap (ODT). Here, $N$ is the field-dressed rotational quantum number and $m_N$ is its projection onto the electric field $\boldsymbol{E}$. We prepare the molecular gases following the procedure described in Ref.[35]. Briefly, we start with a degenerate mixture of $^{40}$K and $^{87}$Rb in an ODT. Molecules in $|0,0\rangle$ are created via magnetoassociation around 546.62 G followed by stimulated Raman adiabatic passage (STIRAP) at an electric field of $\boldsymbol{E}_{STIRAP}$ = 4.5 kV/cm. The molecules are transferred to $|1,0\rangle$ with a microwave Rabi $\pi$–pulse. For detection, molecules are transferred back to $|0,0\rangle$, dissociated to the Feshbach state, and imaged after time-of-flight expansion. The ODT has trapping frequencies of ($\omega_x$, $\omega_y$, $\omega_z$) = $2\pi \times$ (45, 250, 40) Hz for $|1,0\rangle$ at $\boldsymbol{E}_{STIRAP}$, with weak dependence on $\boldsymbol{E}$ (Methods). In contrast to Ref.[42,44] where only the lowest harmonic level along the tightly confined direction is dominantly populated, here we have $k_B T \gg \hbar\omega$ for all three directions, fulfilling the criterion of a 3D geometry ($\hbar$ is the reduced Planck constant and $k_B$ is the Boltzmann constant).

## LONG-LIVED GASES OF POLAR MOLECULES IN 3D

The two-body loss of KRb in $|N = 1, m_N = 0\rangle$ is suppressed at certain electric fields by resonant collisional shielding[44–46]. The suppression arises from tuning collisional channels into degeneracy using $\boldsymbol{E}$, where they are mixed by the resonant dipolar coupling. The strength of the coupling depends on the spatial separation $R$ between the two molecules. The resultant avoided crossing modifies the energy of the coupled channels and manifests as an effective intermolecular potential (see Supplementary Section 1). At the center of the resonance, $|\boldsymbol{E}_0| \approx 12.67$ kV/cm, the energy of the collision channel $|1,0\rangle$ $|1,0\rangle$ is degenerate with that of $|0,0\rangle$ $|2,0\rangle$, where the two kets represent the symmetrized rotational states of the pair of colliding molecules. In the vicinity of $|\boldsymbol{E}_0|$, for $|\boldsymbol{E}| > |\boldsymbol{E}_0|$, the mixing results in a repulsive energy barrier for $|1,0\rangle$ $|1,0\rangle$, preventing the molecules from getting close enough to undergo a chemical reaction. In contrast to the barrier formed by the direct dipolar interaction[26,37,40–42], the barrier formed by resonant shielding exists for both "head-to-tail" and "side-by-side" collisions, shown in Fig. 1**a**, enabling the suppression of two-body loss even in 3D. For $|\boldsymbol{E}| < |\boldsymbol{E}_0|$, the effective potential is mainly attractive, resulting in enhanced loss (Fig. 1**b**).

We measure the shielding effect by monitoring the decay rate of the average molecular density $n$ at different $|\boldsymbol{E}|$. To do so, $\boldsymbol{E}$ is ramped from $\boldsymbol{E}_{\text{STIRAP}}$ to its target value in 60 ms after molecules in $|1,0\rangle$ are produced. We typically have $1.5 \times 10^4$ molecules in $|1,0\rangle$ at $T = 300$ nK with $n = 2.5 \times 10^{11}$ cm$^{-3}$ after the field ramp. After a variable hold time $t$, $\boldsymbol{E}$ is

ramped back to $E_{STIRAP}$ for imaging. The reactive two-body loss rate $\beta$ is extracted by fitting the decay of $n$ with $\dot{n} = -\beta\, n^2 - (3n/2T)\dot{T}$ , where the first term is the reactive loss and the second term accounts for the temperature dependence of $n$. The resonant shielding effect manifests as two sharp features in $\beta$ around $|E|$ = 11.5 kV/cm and 12.5 kV/cm, as shown in Fig. 2**a**. The two resonant features correspond to the coupled channels $|1,0\rangle\, |1,0\rangle \rightarrow |0,0\rangle\, |2,\pm 1\rangle$ and $|1,0\rangle\, |1,0\rangle \rightarrow |0,0\rangle\, |2,0\rangle$ , respectively. The width of each feature is around 100 V/cm, determined by the differential dipole moments of the rotational states involved and the strength of the dipolar coupling (see Supplementary Section 1). Near the resonances, $\beta$ varies by 3 orders of magnitude within a change of $|E|$ of 0.25 kV/cm. Two decay curves exemplifying this contrast are displayed in Fig. 2**b** and Fig. 2**c**. At $|E_S|$= 12.72 kV/cm, we observe long-lived (~10s) molecular gases in 3D (Fig. 2**b**). When $|E|$ is tuned far from the resonances, $\beta$ increases with $d$, similar to KRb in $|0,0\rangle$ (Ref.[26]). At $|E|$ = 4.5 kV/cm, where the molecules have a similar dipole moment as at $|E_S|$, the lifetime is much shorter (~1s), highlighting the prominent effect of resonant dipolar shielding on the two-body loss.

**ELASTIC COLLISIONS BETWEEN MOLECULES IN 3D**

This long-lived molecular gas offers a practical platform to explore the effect of dipolar elastic collisions between reactive molecules. The elastic collisions occur at the rate $n\sigma_{el}v$, where $\sigma_{el}$ is the elastic cross section and $v$ is the ensemble-averaged relative collisional velocity of two molecules (see Methods). Neglecting the effects of resonant shielding, the molecule-molecule elastic cross section $\sigma_{el}$ for indistinguishable fermionic molecules is predicted to vary only with $d$ in the ultracold regime, approaching a universal value $\sigma_{el} = (32\pi/15)\, a_d^2$ (Ref. [38]). Here, $a_d = (m/2) d^2/(4\pi\varepsilon_0 \hbar^2)$ and $m$ is the mass of the molecule. At $|E_S|$, KRb in $|1,0\rangle$ has $d = -0.08$ D and the universal theory predicts $\sigma_{el} = 2.8 \times 10^{-12}$ cm$^2$.

The resonant dipolar coupling modifies the effective intermolecular potential and hence also modifies the properties of the elastic collisions. The effects depend not only on $|E|$ but also on the statistics of the molecules, which affects the short-range interactions. For fermionic $^{40}$K$^{87}$Rb, this effect is manifested as a sharp enhancement in $\sigma_{el}$ as the resonance is approached from lower $|E|$, though $\gamma$ remains low due to enhanced loss. At $|E_S|$, where the loss is suppressed, $\sigma_{el}$ is predicted to deviate only slightly from the universal value[45]. Calculations[45] yield $\gamma = 17.8$ at $T = 330$ nK (Fig. 1**c**), large enough for thermalization within the ensemble lifetime.

We experimentally demonstrate and characterize the elastic collisions through cross-dimensional thermalization, with a geometry shown in Fig. 3**c**. The molecular gas is heated along the more tightly confining $y$-direction. Elastic collisions redistribute the excess kinetic energy from $\boldsymbol{y}$ to $\boldsymbol{x}$ and $\boldsymbol{z}$. The rate $\Gamma_{th}$ of this relaxation process is proportional to the elastic collision rate as

$$\Gamma_{th} = \frac{n\sigma_{el}v}{N_{coll}}.$$

(1)

Here the factor $N_{coll}$ is physically interpreted as the number of collisions required to thermalize. $N_{coll}$ has been calculated to be 2.7 and 4.1 for $s$-wave[49] and $p$-wave[50] collisions, respectively. For dipolar elastic collisions, theoretical calculations[47] and experiments with magnetic atoms[51,52] have shown that $N_{coll}$ depends on the angle $\theta$ between the dipole and the direction of heating $y$. This is a direct result of the anisotropic dipolar collisions.

We observe the anisotropic elastic relaxation process using the following experimental procedure. The molecules are initially prepared in thermal equilibrium in $|1,0\rangle$ at $|\boldsymbol{E}|$ = 12.72 kV/cm with $\theta = 0°$. The field is then rotated to the target angle $\theta$ in 60 ms. Next, a temperature imbalance between the trap axes is introduced by parametrically heating the molecular cloud along $\boldsymbol{y}$ for 50–100 ms (inset of Fig. 3**c**), which is much shorter than the time scale of thermalization. We create an initial condition of $T_y \approx 2.5\ T_x$. The relaxation process is observed by monitoring the time evolution of $T_y$ and $T_x$ as the sample thermalizes in the trap.

As the gas equilibrates, the temperatures approach each other. The temperature evolution for $\theta = 45°$ and $\theta = 90°$ is shown in Fig. 3**a** and 3**b**, respectively. A clear difference in the thermalization rates for the two orientation angles is observed. We quantitatively study the relaxation by fitting the time evolution of $T_x$, $T_y$, and $n$ with a set of coupled differential equations (Methods). The equations capture two physical processes contributing to the temperature change: the elastic dipolar collisions associated with $N_{coll}$, and the reactive loss, associated with the loss coefficient $K_L$, which preferentially removes two colliders with high relative kinetic energy[26]. Here, $K_L$ is related to $\beta$ as $K_L = (1/3)\,\beta/T$ for a thermally equilibrated gas at temperature $T$.

For each $\theta$, we extract $N_{coll}$ and $K_L$ from the fits (Methods). Since the thermalization rate depends on $\sigma_{el}/N_{coll}$, we use $\sigma_{el} = 2.8\times10^{-12}\ \text{cm}^2$ in our analysis to extract $N_{coll}$. We observe a clear angular dependence of the number of collisions required for thermalization, summarized in Fig. 3**c**. At 45°, only $1.6^{+0.2}_{-0.1}$ collisions are required for thermalization while $4.1^{+0.9}_{-0.6}$ collisions are required for $\theta = 90°$. In the limit of small parametric excitation and using the scattering cross section of point dipoles, $N_{coll}$ can be calculated analytically within the Enskog formalism[53]. Within this formalism, the collective behavior of a thermal gas is studied by observing only the phase-space-averaged values of relevant quantities (e.g. the width of the gas cloud) over time, offering a simplification over tracking all molecular degrees of freedom. Adopting the computational techniques used in Ref.[53,54] permits the concise analytic expression:

$$N_{\text{coll}}(\theta) \approx \frac{97.4}{45 + 4\cos 2\theta - 17\cos 4\theta} \tag{2}$$

for a gas heated along **y**, and rethermalization measured along **x**. More details of the derivation are provided in the Supplementary Section 2. Equation (2) (gray solid curve, Fig. 3**c**) describes the measured angular dependence quite well, despite the approximations above. From the measured $K_L$, we calculate $\gamma = \sigma_{el}\,v/(3K_L T)$ as high as 12(1), confirming the dominant role of the elastic collisions in the observed temperature evolution. Our measurements now establish the fact that characteristics of the direct dipolar interaction are preserved at |$E_s$|, suggesting that molecules at the shielding field provide a general platform for realizing stable and strongly dipolar systems.

## EVAPORATIVE COOLING OF MOLECULES IN 3D

A large $\gamma$ enables direct evaporative cooling of KRb in 3D. Moreover, the Wigner threshold law[25,55] suggests $\gamma$ will increase further at lower temperatures for fermionic molecules as $\gamma \sim 1/\sqrt{T}$, which facilitates the evaporative cooling processes. We perform evaporation by lowering the depth of the optical trap at $\theta = 0°$. During evaporation, we observe that the $x$ and $y$ directions remain in equilibrium and the temperature $T$ drops along with the number of molecules $N_{KRb}$ remaining in the trap. Efficient evaporation requires the slope $S_{evap} = \partial \ln N_{KRb} / \partial \ln T$ to be smaller than 3 in 3D. We measure $S_{evap} =$ 1.84(9), significantly below this threshold (Fig. 4). As $N_{KRb}$ decreases, the PSD increases from 0.014(1) to 0.06(2), corresponding to a decrease of $T/T_F$ from 2.3(1) to 1.4(2) ($T_F$ is the Fermi temperature).

Compared with the procedure in Ref.[35] that produced a degenerate Fermi gas at $T/T_F =$ 0.3 with $N_{KRb} = 2.5 \times 10^4$, the present approach requires preparing molecules in |1,0⟩ at |$E_s$|. This requires a ramp of the electric field that causes molecular loss and heating, limiting the highest PSD achieved in this work. Future technical improvements, such as direct creation of molecules at |$E_s$|, will enable evaporation of molecular gases to deep quantum degeneracy. The efficiency of the evaporative cooling demonstrated here is limited by the relatively small $\gamma$ achievable in KRb. For molecules such as $^{23}Na^{133}Cs$, the same approach should allow $\gamma > 10^6$ at moderate electric fields around 2.5 kV/cm (Ref [48]), enabling more efficient evaporation.

**CONCLUSIONS**

Employing an electric field-tuned shielding resonance, we have demonstrated anisotropic thermalization via dipolar elastic collisions and performed efficient evaporative cooling of reactive polar molecules in 3D. The two-body loss is greatly suppressed by resonant dipolar shielding, giving a ratio of elastic to inelastic collision rates as high as 12. Our work highlights a general approach for controlling the interaction properties of polar molecules in 3D. The same mechanism has been predicted to be effective for other molecules, for which even higher ratios [48] and even more efficient evaporation may be achievable. Our findings demonstrate a promising strategy for producing low-entropy molecular gases in bulk systems and open the door for a broad range of applications in molecule-based quantum platforms.

*Note: During the preparation of this manuscript we became aware of a recent work (Ref.[56]) reporting microwave shielding in a molecular tweezer.*

**METHODS**

**Thermalization Model**

We fit the temperature and density evolution with a set of differential equations[26]:

$$\dot{n} = -K_L(T_y + 2T_x)n^2 - \frac{n}{2T_y}\dot{T_y} - \frac{n}{2T_x}\dot{T_x}$$

$$\dot{T_y} = \frac{n}{4}K_L(-T_y + 2T_x)T_y - \frac{2\Gamma_{th}}{3}(T_y - T_x) + c_y$$

$$\dot{T_x} = \frac{n}{4}K_L T_y T_x + \frac{\Gamma_{th}}{3}(T_y - T_x) + c_x. \qquad (3)$$

$\Gamma_{th}$ is defined in equation (1) with $v = \sqrt{16k_B(T_y + 2T_x)/(3\pi m)}$. $K_L$ describes the two-body loss and $c_y$ and $c_x$ are background heating rates. The model captures the two main contributions to the observed temperature evolution: the elastic dipolar collisions described by the term proportional to $\Gamma_{th}$, and the effects on the temperature from the reactive loss described by the first term related to $K_L$.

The model has two assumptions for simplicity. First, we assume a similar reaction coefficient $K_L$ for molecules colliding along different directions with respect to the dipole, which is valid in the vicinity of |**E**s|[44]. Second, the temperatures of the two unmodulated directions $x$ and $z$ remain identical during the thermalization process. Systematic error introduced by this assumption is maximized at the angle of 45° where the thermalization speed between **y**, **x** and **y**, **z** differs by the most. Since the thermalization between **y**, **z** is much slower, adopting this assumption leads us to underestimate the thermalization rate around 45°.

**Data Analysis**

For each $\theta$, we image the molecules at several hold times between 0.05 and 10 s, at several times-of-flight between 1.5 and 8.2 ms, and with and without parametric heating. To minimize the systematic effects from slow drift in experimental conditions, we randomize the order in which the data is taken. For each hold time and heating condition, we fit the temperatures $T_x$ and $T_y$ and average density $n$, assuming free expansion of the cloud. Estimated values and their covariance matrices are obtained via bootstrapping. We fit the temperature and density decay curves to equation (3), treating $K_L$, $N_{coll}$, $c_x$, $c_y$, and the initial temperatures and densities as fit parameters. To estimate confidence intervals on the fit parameters, we generate 100 synthetic datasets by independently drawing new temperatures and densities at each time point from a multivariate normal distribution. The reported parameters and confidence intervals represent the median and the intervals containing 68% of the trials. This approach lets us examine correlations between the different fit parameters. An example fit is shown in Extended Data Fig.1**a** for $\theta = 45°$. The fit yields $N_{coll} = 1.57(14)$, $K_L = 3.8(3) \times 10^{-7}$ cm$^3$s$^{-1}$K$^{-1}$.

Extended Data Fig. 1**b** shows the fitted result for 100 synthetic datasets. The small correlation of -0.26 between $N_{coll}$ and $K_L$ indicates that the fitting distinguishes between the thermalization and two-body loss well.

Extended Data Fig. 1**c** shows the extracted loss rate $K_L$ vs. $\theta$. Though the measured $K_L$ seems to be anti-correlated with $N_{coll}$, we find that the observed variation of $K_L$ is not drastic enough to cause a significant change to the extracted $N_{coll}$. This weak modulation of $K_L$ could be caused by daily technical drifts on the electric field strength or residual field gradient, or contributions of the higher partial wave in the two-body collisions which may cause a higher $K_L$ around $\theta \approx 54°$ where the shielding barrier is weak.

**Temperature Measurement**

Temperatures are measured by fitting the Gaussian width of the cloud after time-of-flight expansion. To image the molecules, the electric field $\boldsymbol{E}$ is ramped from $\boldsymbol{E}_S$ back to $\boldsymbol{E}_{\text{STIRAP}}$ after the hold time. The molecules are then transferred to $|0,0\rangle$ and dissociated for imaging. The changes in the bias field and molecular rotational state result in changes to the molecular polarizability and, consequently, the trapping potential. To obtain accurate temperatures of the molecular gas during the thermalization process, we correct for such systematics as detailed below.

Transferring from $|1,0\rangle$ to $|0,0\rangle$ while the molecules are still in the trap causes an instantaneous change of the trapping potential and thus breathing of the cloud, which would introduce errors in the measured temperatures. We avoid this issue by performing the rotational state transfer and STIRAP during time-of-flight.

The ramp down of the electric field from the measurement condition to $\boldsymbol{E}_{\text{STIRAP}}$ modifies the trapping potential due to the polarizability change of $|1,0\rangle$ molecules with the bias field. Since this ramp is adiabatic with respect to the single particle trapping periods along all the spatial direction, the resultant adiabatic compression/decompression modifies the molecular temperature. We calculate the actual temperatures $T_i$ during the thermalization process from the measured temperature $T_i^{\mathrm{m}}$ by

$$T_i = \left( \frac{\omega_i^S}{\omega_i^{\text{STIRAP}}} \right) T_i^{\mathrm{m}}. \tag{4}$$

Here, $\omega_i^S$ and $\omega_i^{STIRAP}$ are the trapping frequencies along $i$ at $E_S$ and $E_{STIRAP}$, respectively. The trapping frequencies are calibrated by measuring the parametric heating resonances at each theta. The temperatures reported in the manuscript are $T_i$.

**Data Availability**

The data that support the findings of this study are available from the corresponding author upon reasonable request. Source data are provided with this paper.

**ACKNOWLEDGEMENTS**

We acknowledge funding from ARO-MURI, AFOSR-MURI, DARPA DRINQS, NSFQLCI OMA–2016244, NIST, and NSF grant 1806971. L. L. and G. Q. acknowledge funding from the FEW2MANY-SHIELD Project No. ANR-17-CE30-0015 from Agence Nationale de la Recherche. We thank Lee R. Liu for reading the manuscript.

**AUTHOR CONTRIBUTIONS**

J.-R. L., W. G. T., K. M., C. M., G. V., L. D. M., and J. Y. contributed to the experimental measurements. L. L. and G. Q. calculated the effective intermolecular potential. R. R. W. W. and J. L. B. calculated the anisotropic thermalization rate. All authors discussed the results, contributed to the data analysis, and worked on the manuscript.

**COMPETING INTERESTS**

The authors declare no competing interests.

**MATERIALS & CORRESPONDENCE**

Correspondence and material requests should be addressed to Jun-Ru Li (junru.li@colorado.edu) and Jun Ye (Ye@jila.colorado.edu).

**AUTHOR INFORMATION**

Giacomo Valtolina:

Present address: Fritz-Haber-Institut der Max-Planck-Gesellschaft, Faradayweg 4-6, 14195Berlin, Germany

**ADDITIONAL INFORMATION**

**Supplementary Information** section 1, 2 including Fig.S1 are available for this paper.

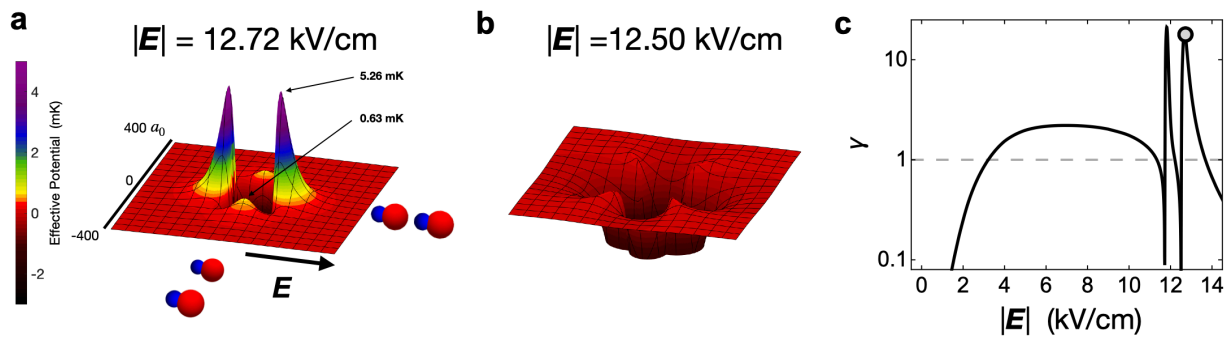

**Figure 1. Effective intermolecular potential and tuning of molecular interactions near the shielding resonance. a** and **b**, Calculated effective intermolecular potential for KRb in |1,0⟩ at |$\textbf{E}$| = 12.72 kV/cm and |$\textbf{E}$| = 12.50 kV/cm respectively (see Supplementary Section 1). The resonant dipolar coupling mixes two degenerate collisional channels. The strong mixing between the channels modifies the effective intermolecular potential and results in either suppression or enhancement of the two-body loss rate depending on |$\textbf{E}$|. $a_0$ is the Bohr radius. **c**, Calculated ratio $\gamma$ of the elastic rate to reactive rate for KRb in |1,0⟩ at $T$ = 330 nK[45]. Our experiment is carried out at |$\textbf{E}_s$| = 12.72 kV/cm (indicated by the gray point) where $\gamma$ of 17.8 is calculated. The gray dashed line indicates $\gamma$ = 1.

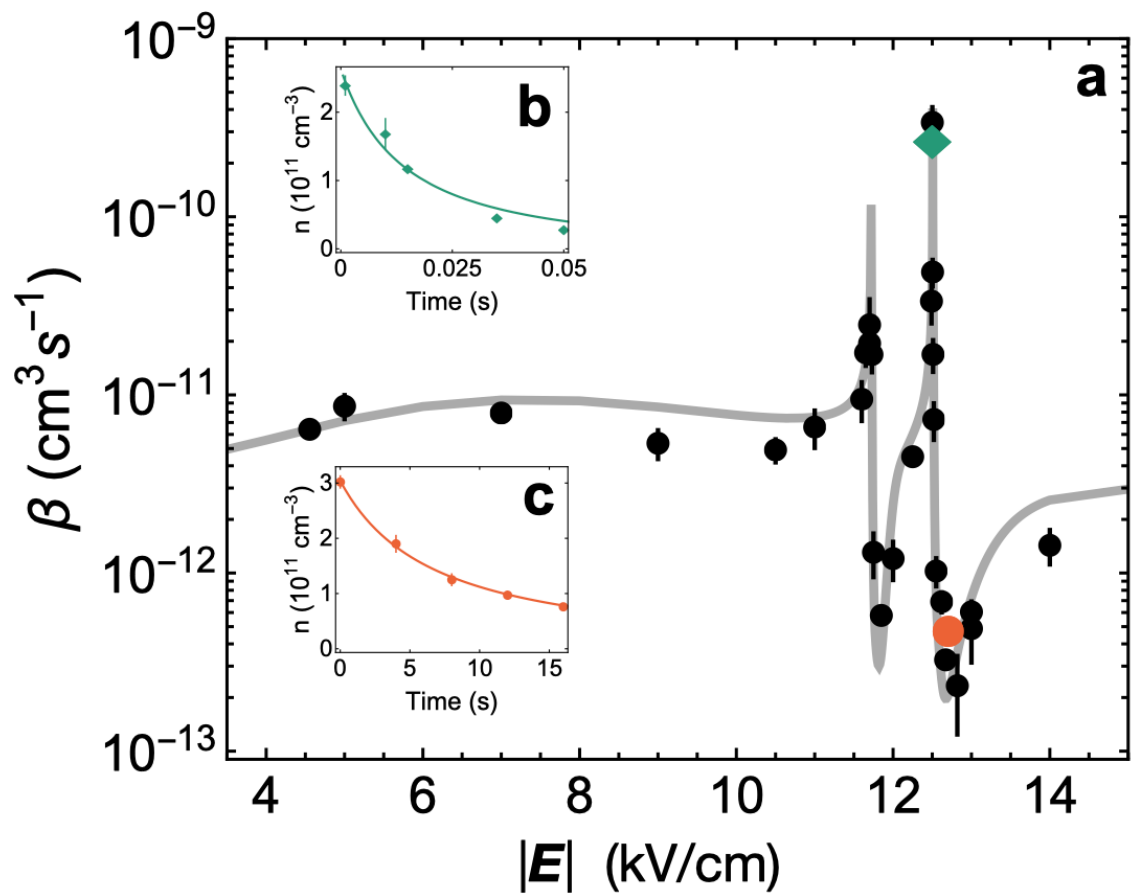

**Figure 2. Resonant shielding of the reactive loss in 3D. a,** Measured two-body loss rate versus electric field strength. The solid line is the theoretical calculation for the experimental condition $T$ = 330 nK with no free parameters. The green diamond and orange circle identify the fields for which decay curves are plotted in insets **b** and **c**, respectively. Insets: Molecule loss measurements at |$E$| = 12.50 kV/cm (**b**) and 12.72 kV/cm (**c**) where the loss is enhanced and suppressed respectively. Solid lines are fits to the two-body loss rate equation.

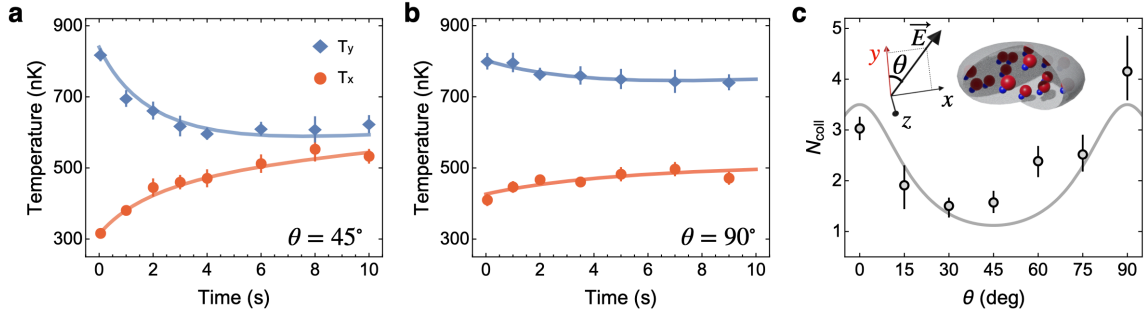

**Figure 3. Anisotropic cross-dimensional thermalization of molecules via dipolar elastic collisions. a** and **b**, Time evolution of the temperature for $\theta$ = 45° and $\theta$ = 90° after parametric heating along **y**. The solid lines are fits to our model (Methods). The thermalization is faster at $\theta$ = 45°. $T_z$ is not directly measured and is assumed to be the same as $T_x$ during the entire process. Error bars are 1 SE. **c**, Angle-dependent number of collisions required for the rethermalization of dipolar Fermi gases. $N_{coll}$ is extracted by fitting the time evolution of the temperature (Methods). The gray solid line represents the calculated analytical expression, equation (2). Inset: Geometry of the experiment. Molecules are polarized with a bias field **E** whose orientation angle $\theta$ is varied between 0° and 90° within the x–y plane. The molecular gases are heated parametrically along the y direction to create an initial condition of $T_y \approx 2.5\ T_x$, $T_x = T_z$.

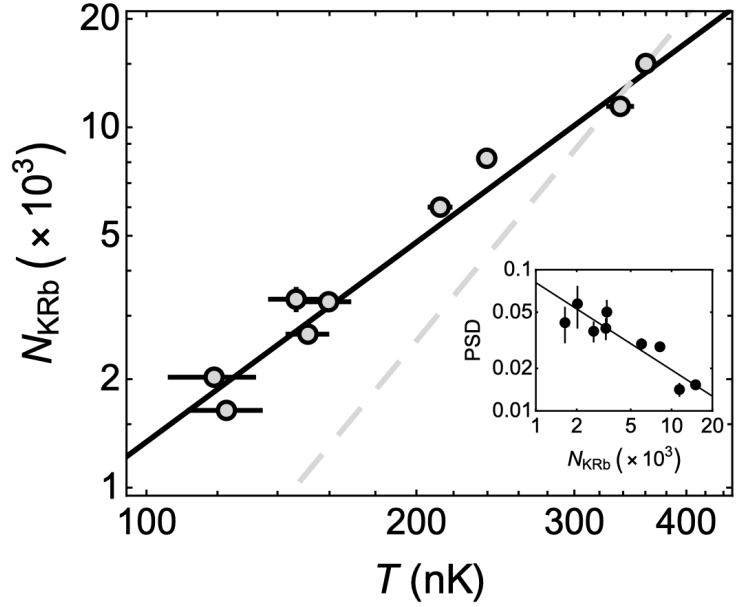

**Figure 4. Efficient evaporative cooling of reactive polar molecules in 3D.** Evolution of $N_{KRb}$ and $T$ at different stages of the evaporation at |$\textbf{\textit{E}}_s$| and with $\theta = 0°$. The power-law fits (black line) yields $S_{evap} = 1.84(9)$, indicating efficient evaporation. The dashed line represents a constant PSD. Error bars are 1 SE. Inset: PSD $(N_{KRb}(\hbar\bar{\omega}/k_B T)^3)$ versus $N_{KRb}$ during evaporation, displaying a clear gain. Here, $\bar{\omega}$ is the geometric mean of the trapping frequencies.

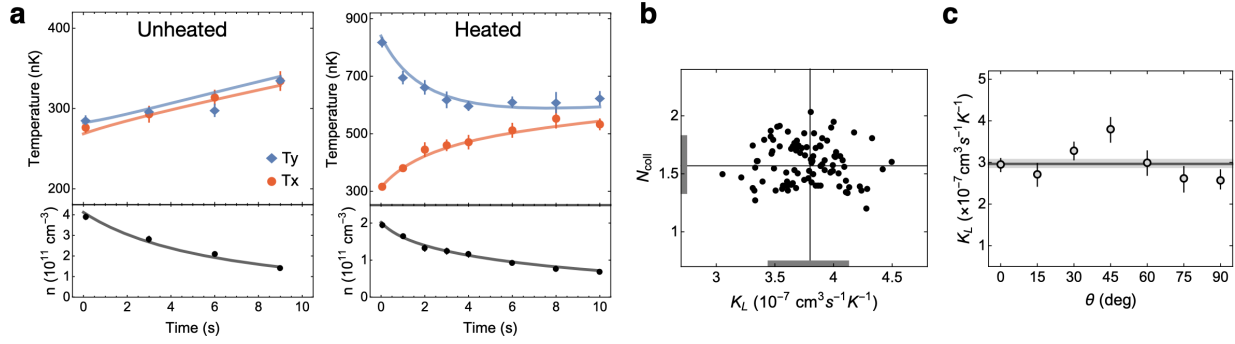

**Extended Data Figure 1.** Fitting experimental data with the model. The figure shows the fit results for $\theta = 45°$. **a**, Fitting of the unheated and heated data. **b**, Fitted $K_L$ and $N_{coll}$ for 100 synthetic datasets. We extract a correlation of -0.26 between the two fitted parameters, indicating that the fitting can distinguish between two-body loss and thermalization. The black solid lines are the median of all the fitted results from the synthetic datasets for $N_{coll}$ and $K_L$, while the gray lines on the axis represent 68% confidence interval of the fitted results. This median and 68% confidence are reported in the main text. **c**, Extracted loss coefficient $K_L$ versus $\theta$. The line and shaded region indicate the mean value of $K_L$ and its standard error.